\documentclass[10pt,conference]{IEEEtran}
\IEEEoverridecommandlockouts
\usepackage{cite}
\usepackage{amsmath,amssymb,amsfonts}
\usepackage{algorithmic}
\usepackage{graphicx}
\usepackage{textcomp}
\usepackage{xcolor}
\usepackage[utf8]{inputenc}
\usepackage{url}
\usepackage{soul}

\def\BibTeX{{\rm B\kern-.05em{\sc i\kern-.025em b}\kern-.08em
    T\kern-.1667em\lower.7ex\hbox{E}\kern-.125emX}}

\begin{document}

\title{A Methodology for Using GitLab for Software Engineering Learning Analytics}

\author{\IEEEauthorblockN{Julio C\'esar Cort\'es R\'ios, Kamilla Kopec-Harding, Sukru Eraslan, \\
Christopher Page, Robert Haines, Caroline Jay and Suzanne M. Embury} \\
\IEEEauthorblockA{\textit{University of Manchester}\\
Manchester, United Kingdom \\
juliocesar.cortesrios@manchester.ac.uk}}

\maketitle

\begin{abstract}
To bridge the digital skills gap, we need to train more people in Software Engineering techniques. This paper reports on a project exploring the way students solve tasks using collaborative development platforms and version control systems, such as GitLab, to find patterns and evaluation metrics that can be used to improve the course content and reflect on the most common issues the students are facing. In this paper, we explore Learning Analytics approaches that can be used with GitLab and similar tools, and discuss the challenges raised when applying those approaches in Software Engineering Education, with the objective of building a pipeline that supports the full Learning Analytics cycle, from data extraction to data analysis. We focus in particular on the data anonymisation step of the proposed pipeline to explore the available alternatives to satisfy the data protection requirements when handling personal information in academic environments for research purposes.
\end{abstract}


\section{Introduction}
\label{sec:introduction}

The world is currently in the grip of a chronic digital skills shortage. A number of initiatives are trying to close the gap between supply and demand~\cite{Royle2014, vanDijk2014, Morel2013, Varallyai2013}. Among these, the UK Institute of Coding (IoC)~\cite{Davenport2019}\footnote{\url{instituteofcoding.org}} is exploring how to provide better training in this area. Here we report on a project exploring how data on novice software engineers' usage of version control and continuous integration systems can be used to understand the way people learn to program, and ultimately to create feedback and marking tools to support them in this endeavour. This learning improvement based on material collected from the educational process is called Learning Analytics (LA)~\cite{siemens2011, ferguson2012}, a framework that can help in bridging the gap in digital skills education.

The ease or difficulty of acquiring new skills is an important human factor in software engineering.  While a great deal of research has been carried out into how humans learn to code, much less attention has been given to how we acquire other key software engineering skills, and how to make that learning more efficient and effective. With this aim in mind, we are analysing artefacts produced by students on our Software Engineering (SE) course units at the University of Manchester, UK, to better understand the pitfalls and challenges of learning core SE skills. In these courses, students learn how to perform collaborative development and version control, with the objective of preparing them for real-world SE projects. Ensuring that research data, and in this instance student data, is collected and used ethically is of paramount importance. As a first step, we are constructing and evaluating the process of data extraction and handling, to satisfy data protection policies, and minimise the risk of participant re-identification.

Taking into account only the mechanism needed to extract and prepare the information for the analysis is not sufficient when dealing with data generated as part of an educational environment. Further challenges arise from the fact that learning digital skills, such as the use of a distributed version control system like Git, or a Git Repository Manager such as GitLab, require understanding two things: is the student learning about, and exploiting, the tool's capabilities, and; how the interaction with these tools changes at different stages of the learning process. For example, how the student is collaborating with other students, within the same team, using metrics generated from their repository, such as the entropy computation based on their commits~\cite{Mittal:2014:PMS:2591062.2591152}. A few works have addressed these challenges of teaching and learning SE topics. For instance, Isom{\"o}tt{\"o}nen and Cochez explore the problems arising when learning Git from the students' perspective, and how these problems are aligned with the stage of the learning the students are currently in, and how incomplete the assimilation is if the focus is on how the students interact with the tools instead of understanding how they are assimilating the concepts~\cite{Isomottonen2014}. Haaranen and Lehtinen explored the user interaction with Git from the instructor and learner perspectives~\cite{Haaranen2015}, and found that just teaching Git concepts is insufficient, if they not accompanied by a practical (and appealing) scenario in which the students can make mistakes and experiment at their own pace.

As seen in previous works, in the case of SE skills, such as Git and GitLab, it is important to apply the basic concepts in practical situations where the students can openly use these tools and improve their understanding, as explored in~\cite{wohlin2012}. An SE practice that follows a similar approach is Test Driven Development (TDD)~\cite{beck2003, lee2017, suleman2017}, in which students learn by testing first over small code units, to detect issues on specific requirements, and then re-factoring the code to satisfy these requirements, with the aim of producing just essential code and reduce the steps to get a functional version of the system. In many cases, TDD is applied into educational games projects~\cite{caulfield2011}, and the objective is similar to the one pursued by our research: to consider the stages of the learning experience whilst supporting an adequate and gradual application of the course concepts based on the students' assimilation rate.

The previous examples show that it is not enough to conduct the analysis of the performance of students learning SE skills by collecting metrics on the platforms used. We also need to consider the human aspect of gradually introducing practical scenarios that can be tackled at varied speeds. These differences in the learning process must be inspected when evaluating the students and also when analysing further improvements, applying LA, to the teaching and learning process.

In this paper, we summarise how LA methods have previously been used in SE education, with a focus on how student data generated through interaction with GitLab can be handled to gather significant information not only about the system-user interactions but also about the mistakes and lessons learned along the way. Then we present a description of a proposed pipeline for LA that incorporates the steps needed to prepare the data for LA, from data extraction to data analysis, and taking into account the need for confidentiality, for which an exploration of alternatives for data anonymisation was performed. Finally, conclusions are presented.
\section{Learning analytics approaches}
\label{sec:approaches}

The use of data extracted from SE courses for learning purposes has been explored before, but the focus has generally been on \textit{ad-hoc} tools and mechanisms that support data collection, and without taking into account the human aspects of the information that is being collected. For instance, a student might make mistakes due to a lack of understanding of how the merging mechanism works in Git, hence, the analysis of error patterns along the course time line could potentially support a larger improvement for that particular student than just determining if the course objectives were satisfied or not. Works such as~\cite{conde2014, Conde2015, martinez2015, Echeverria2016} rely on e-Learning platforms to collect interactions between students and the system, and apply visual learning analytics to detect areas of improvement based exclusively on those interactions. 

Other research has used data mining to collect information about the competencies of students to determine the quality of the educational experience for business purposes~\cite{Misnevs2018}, and examined data from Massive Open Online Courses (MOOCs) to determine gaps in the provision of SE skills~\cite{DeOliveira2017}. These approaches rely on an existing integration between the students' learning platform and the tool or service supporting the LA, which is convenient if such platforms are available. However, if such an integration does not exist, or there is no definition about what information would be required to perform the LA, there is no guarantee that the learning platform will be able to provide what is needed.

In a similar situation to the one presented in Section~\ref{sec:introduction}, P\'erez-Berenguer and Garc\'ia-Molina~\cite{PerezBerenguer2018} and 
Perez-Colado \textit{et al.}~\cite{PerezColado2018} use didactical games to collect the information needed by the LA process but, instead of tailoring this integration between LA and the gaming platform, they separate the LA process into two parts: one using a Learning Analytics Model (LAM) to describe the analysis in terms of the learning exclusively; and an independent analytics system to interact with the gaming platform, which focuses on other implementation aspects of the user experience such as security, flexibility and performance. The latter approaches provide a platform-independent solution that is relevant to our scenario.

The research presented in this paper is therefore based on the latter approaches. We use data logged from the GitLab version control system as a starting point for understanding learner behaviour, and complement this information with data gathered directly from Git repositories and Continuous Integration systems. We compare the actual achievement against the course objectives and evaluate the teams' performance, not only against the expected outcomes, but also by analysing the communication that took place during the process and the most common mistakes---such as committing code changes to the wrong branch---that are made while the students are completing the exercise objectives.
\section{Data processing pipeline}
\label{sec:pipeline}

In this section, a practical scenario using a pipeline for LA applied to SE is presented. It takes into account the sensitive nature of the data generated by the students during the learning process, while gathering enough information to provide a meaningful view of the learning experience. The data includes the most common mistakes made while learning, and how these mistakes were handled by the students.

\begin{figure}[!htbp]
      \centering
      \scalebox{0.33}{\includegraphics{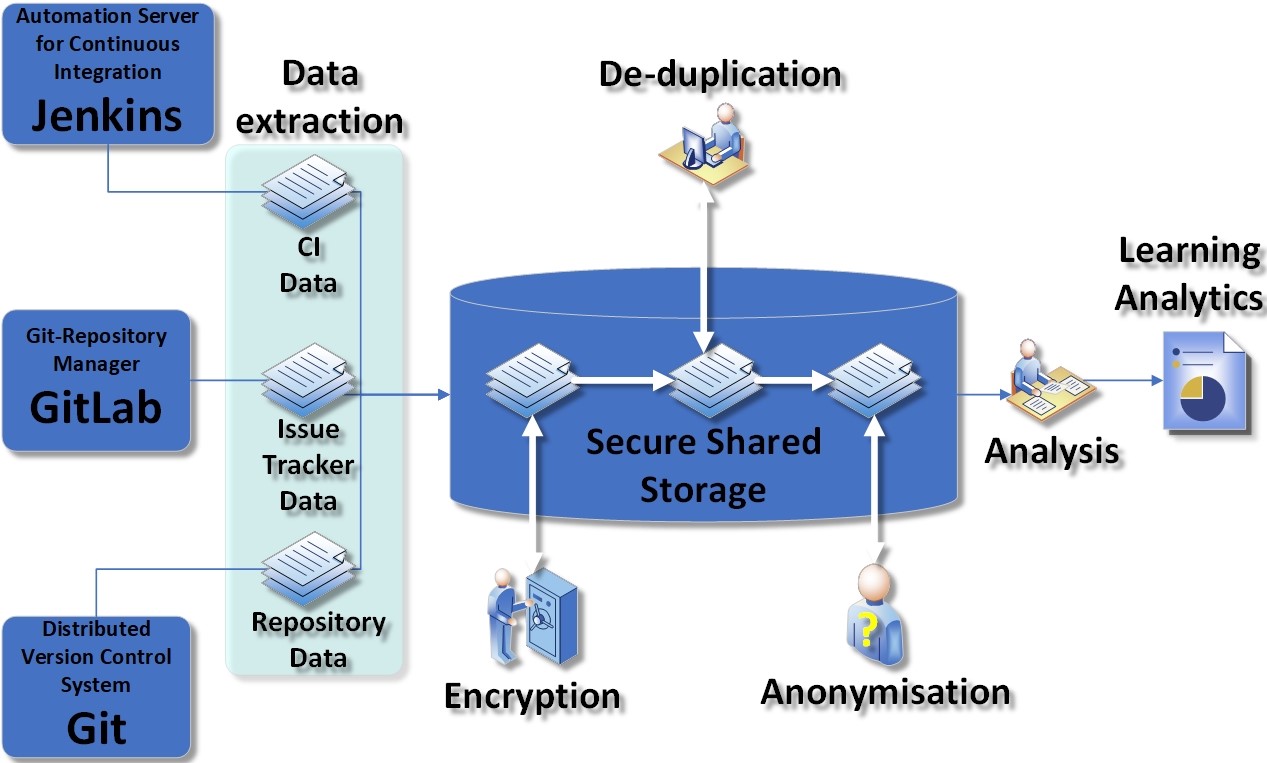}}
      \caption{Data processing pipeline}
      \label{fig:pipeline}
\end{figure}

As shown in Figure~\ref{fig:pipeline}, the pipeline draws data from the learning resources provided to the students. In this scenario these are: a software repository for the student projects in Git; a repository manager (GitLab), that keeps track of the SE cycle during the exercise; and a system for Continuous Integration (Jenkins), that automatically performs software testing and deployment. The combined usage of these tools aims to simulate a real SE environment.

The pipeline uses Jenkins instead of GitLab CI to comply with the requirements of our software engineering (SE) course units at the University of Manchester. The student repositories are privately stored within the University's IT infrastructure.

Once the learning resources are identified and available, the first step in the proposed pipeline is to extract structured data by combining data extraction techniques and direct API integration. The goal is to create a data set containing a description of the students' experience during the learning process. The next step is to ensure that all the information extracted is stored securely: the extracted data sets are encrypted at rest to prevent unauthorised access. Once security is enforced we clean the data, which in this case consists of handling duplicate users found in the Git and GitLab operations, caused by students using several devices while working on the course tasks. The result of this de-duplication is a data set where individuals are fully identified. Next, an anonymisation process removes any data that can be used to identify an individual from the data set. This step is critical to comply with data protection policies concerning students' data. An exploration of techniques that can be used for this step are presented in Section~\ref{sec:anonymisation}. Finally, once the data extracted is secure, clean and private, the analysis can be performed to detect trends and indicators that can be used to understand and improve the learning experience, following the LA approach, as explored in Section~\ref{sec:approaches}.

The proposed pipeline is designed to deal with the challenges of handling student-generated data for Learning Analytics purposes, and, as part of our research, we aim to fully implement all proposed steps and apply the pipeline in a scenario with real students' data gathered from SE courses.

\section{Data anonymisation}
\label{sec:anonymisation}

A critical aspect of the proposed pipeline from Section~\ref{sec:pipeline} is to enforce the data protection policies that are in place in academic institutions, with the aim of minimising the risk that students' data is misused. The ethical implications must be taken into account even if the purpose of the research is to improve the educational experience. Data protection policies aim to safeguard participant information from unauthorised processing, and one way to achieve this is to reduce the risk of participant re-identification through anonymisation. In particular, correlation analysis may be required between some metrics and student marks, and the anonymisation process should ensure that this kind of analysis does not reveal sensitive information. The anonymisation also needs to consider the usefulness of the data once the process is completed: if the loss of utility prevents the application of LA to the anonymised data set then important outcomes could be lost.

For data anonymisation, there are several approaches that can be followed, such as generalisation, permutation, perturbation, suppression, anatomisation and their combinations. These approaches provide variable degrees of anonymisation at the expense of substantial loss of information, and their applicability to our scenario was evaluated.

The generalisation approach is based on the replacement of specific values with generic ones, such as replacing all the telephone numbers in a data set with their correspondent area code. An example of applying this approach can be seen in~\cite{Wilczynski2018}, where the generalisation is applied over data before it is sent through the network by creating a virtualisation layer. In this layer, sensitive data is replaced by ranges, to prevent an association with an individual based on specific features of the information. Generalisation could potentially be applied to our scenario as it may anonymise sensitive information such as marks assigned to the students, by defining ranges, or bins, to classify the marks without discriminating specific students. Such an approach would potentially prevent further analysis of the marks at an individual level, but a trade-off is sometimes required to safeguard the confidentiality of the information.

The next approach is based on permutations of the information to avoid identification based on the correlation between the records contained in a data set. An example of the permutation approach is presented in~\cite{Bild2018}, which uses an iterative method to apply data transformations on adjacent records based on a predefined search strategy until a criterion is fulfilled. In our scenario, the problem with such a solution is that the relation between the records needs to be preserved as it provides a critical insight of how the students completed the exercises in GitLab. Therefore, an approach that focuses the anonymisation on such relations cannot be applied into our scenario without loosing critical data for the analysis.

Perturbation is another anonymisation technique that replaces the original values with different ones that cannot be inferred from the non-anonymised data. The altered information is obtained by adding noise, interchanging values or creating \textit{ad-hoc} data to preserve its utility. In~\cite{Eyupoglu2018}, the perturbation approach is used in combination with chaotic functions to generate new values. Such combined approaches satisfy the requirements for our scenario, as they can be used to selectively generate new data to replace the sensitive values that could be used to identify an individual and, at the same time, preserve the relations and utility of the extracted data sets. On the other hand, the trade-off of applying complex functions to improve the anonymisation is, generally, a complicated mechanism to aggregate and analyse such data (e.g., additional computational time and space needed for the analysis).

Suppression removes sensitive values from the data set to preserve its privacy. However, such an approach is generally avoided as it results in a huge loss in utility of the resulting data sets. Hence, it is commonly combined with other techniques to preserve the data utility to some extent. For instance, Deivanai \textit{et al.} propose combining the generalisation or perturbation techniques with suppression to remove specific values based on other attribute values, with respect to how much these attributes influence data classification~\cite{deivanai2011}. In our scenario, there is no need to suppress values to preserve privacy, as there are other techniques, such as perturbation, that can satisfy the anonymisation without loss of utility.

Finally, the anatomisation approach creates groups of sensitive data based on some predefined criteria. By itself such technique would cause loss of important information, thus it is normally combined with other anonymisation techniques. For instance, in~\cite{saeed2018}, it is combined with generalisation and suppression to guarantee data privacy while preserving utility. Such a combined approach can be applied to our scenario, to anonymise data that could potentially be used to identify an individual, such as commit comments provided by students.

Regarding the extraction and handling of the data collected as part of an SE course, the approaches detailed above provide alternatives to enforce data protection policies while minimising data utility loss for LA purposes. From the existing approaches, generalisation, perturbation and anatomisation could be appropriate to our scenario, in which the information generated in GitLab during the SE course will be used to improve the educational process by applying LA.
\section{Conclusion}
\label{sec:conclusion}

GitLab data extracted from Software Engineering courses can be used for Learning Analytics aimed at improving the learning experience. The analysis of this data must take into account the human aspects of the experience, such as gradual experimentation, learning based on mistakes, and the learning abilities of each individual. Furthermore, the data needs to be properly handled during the extraction, analysis and publication, to comply with data protection regulations, while still preserving the details required for the analysis to provide useful results. This research explored ways in which those considerations have been addressed in other works to support data analysis used to improve the content of Software Engineering courses, and identify those alternatives that are most suited for a scenario in which LA is applied to data collected from Git and GitLab, and how data anonymisation can be used to satisfy the data protection policies. This initial work may inspire additional research to be focused on the complex aspects accompanying the use of student-generated information to improve the learning experience of SE courses.

\section*{Acknowledgements}
Work supported by the Institute of Coding which received funding from the Office for Students (OfS), and support from the Higher Education Funding Council for Wales (HEFCW).

\bibliographystyle{IEEEtran}
\bibliography{main}

\end{document}